\documentclass[sigconf]{acmart}
\AtBeginDocument{%
  \providecommand\BibTeX{{%
    \normalfont B\kern-0.5em{\scshape i\kern-0.25em b}\kern-0.8em\TeX}}}

\setcopyright{none}
\renewcommand\footnotetextcopyrightpermission[1]{}

\usepackage{ifthen}
\usepackage{xcolor}
\usepackage{xspace}
\usepackage{balance}

\newboolean{showcomments}

\setboolean{showcomments}{true}

\ifthenelse{\boolean{showcomments}}
  {
  }

\newcommand{\ie}{\emph{i.e.,}\xspace}
\newcommand{\eg}{\emph{e.g.,}\xspace}

\newcommand{\etal}{\emph{et~al.}\xspace}

\newcommand{\figref}[1]{Fig.~\ref{#1}\xspace}

\begin{document}

\title[Enhancing Trace Visualizations for Microservices Performance Analysis]{Enhancing Trace Visualizations for\\Microservices Performance Analysis}

\author{Jessica Leone}
\email{jessica.leone@student.univaq.it}
\orcid{0000-0002-2870-8161}
\affiliation{
  \institution{University of L'Aquila, Italy}
  \country{}
}

\author{Luca Traini}
\email{luca.traini@univaq.it}
\orcid{0000-0003-3676-0645}
\affiliation{
  \institution{University of L'Aquila, Italy}
  \country{}
}

\titlenote{This is a pre-copy-editing, author-produced version of an article accepted for publication in the 6th Workshop on Hot Topics in Cloud Computing Performance (HotCloudPerf 2023). The final authenticated version is available online at: \url{https://doi.org/10.1145/3578245.3584729}.}

\begin{abstract}
Performance analysis of microservices can be a challenging task, as a typical request to these systems involves multiple Remote Procedure Calls (RPC) spanning across independent services and machines. Practitioners primarily rely on distributed tracing tools to closely monitor microservices performance. These tools enable practitioners to trace, collect, and visualize  RPC workflows and associated events in the context of individual end-to-end requests. While effective for analyzing individual end-to-end requests, current distributed tracing visualizations often fall short in providing a comprehensive understanding of the system's overall performance.
To address this limitation, we propose a novel visualization approach that enables aggregate performance analysis of multiple end-to-end requests. Our approach builds on a previously developed technique for comparing structural differences of request pairs and extends it for aggregate performance analysis of sets of requests.
This paper presents our proposal and discusses our preliminary ongoing progress in developing this innovative approach.
\end{abstract}



\maketitle

\section{Introduction}

Microservices have emerged as a paradigm shift in software development, enabling the efficient testing and deployment of new software features and improvements~\cite{newman2015}. Due to their modular nature, microservices are particularly well-suited for today's software market, where the ability to rapidly release software updates and improvements is perceived as a key competitive advantage~\cite{rubin2016}.

However, despite these advantages, adopting microservices also poses significant challenges. A main challenge, for example, is ensuring adequate software performance.
Indeed, a proactive approach to performance assurance (\eg pre-production performance testing~\cite{jiang2015,laaber2018,traini2022c}) is often impractical in these contexts, given the inherent complexity of these systems \cite{sridharan2017, veeraraghavan2016}, and the significant pressure to deliver fast-to-market \cite{rubin2016, traini2022}.
Performance behavior of microservices systems usually \emph{emerges} in the field \cite{veeraraghavan2016} as a complex interaction of several RPCs that spans multiple independent services and machines.
Furthermore, these systems are constantly undergoing software changes (\eg multiple releases per day) and are exposed to highly varying workloads \cite{ardelean2018}, which make them susceptible to unexpected performance regressions~\cite{veeraraghavan2016,traini2021}.

To aid performance analysis, practitioners typically rely on distributed tracing~\cite{parker2020} (\eg Jaeger~\cite{jaeger}, Dapper~\cite{sigelman2010}, Canopy\cite{kaldor2017}) as a means to monitor and inspect microservices performance in production.
Distributed tracing tools capture the workflow of causally-related events (\ie work done to process a request) within and among the components of a microservices system \cite{sambasivan2016}, and provide visualizations to support performance analysis of individual end-to-end requests, \eg swimlane visualizations~\cite{Davidson2023, sigelman2010, jaeger}.

Although these visualizations are particularly effective for analyzing individual requests, they are limited when it comes to provide a comprehensive picture of microservices performance.
Understanding request performance in \emph{aggregate} \cite{parker2020} often requires switching between various visualization tools, making the process cumbersome and time-consuming~\cite{Davidson2023}.
Current distributed tracing tools do not provide out-of-the-box support for aggregate performance analysis.

In this paper, we propose a novel visualization approach that facilitates comprehension of the relationship between request characteristics and end-to-end response time behavior.
The proposed approach enables seamless performance analysis of multiple requests, eliminating the need to switch between multiple tools and visualizations.
Our approach builds upon a previously developed visualization method for comparing request pairs \cite{farro2018} and extends it to accommodate various types of aggregate analysis on sets of requests.
This new approach will empower engineers to swiftly identify significant and recurring request characteristics that are linked to specific end-to-end response time behavior, thereby enabling a more effective performance analysis of microservices.
Here, we present our proposal and report on our preliminary ongoing progress in developing this novel visualization approach.

\section{Motivation}
\begin{figure}[h]
    \centering
    \includegraphics[width=0.8\linewidth]{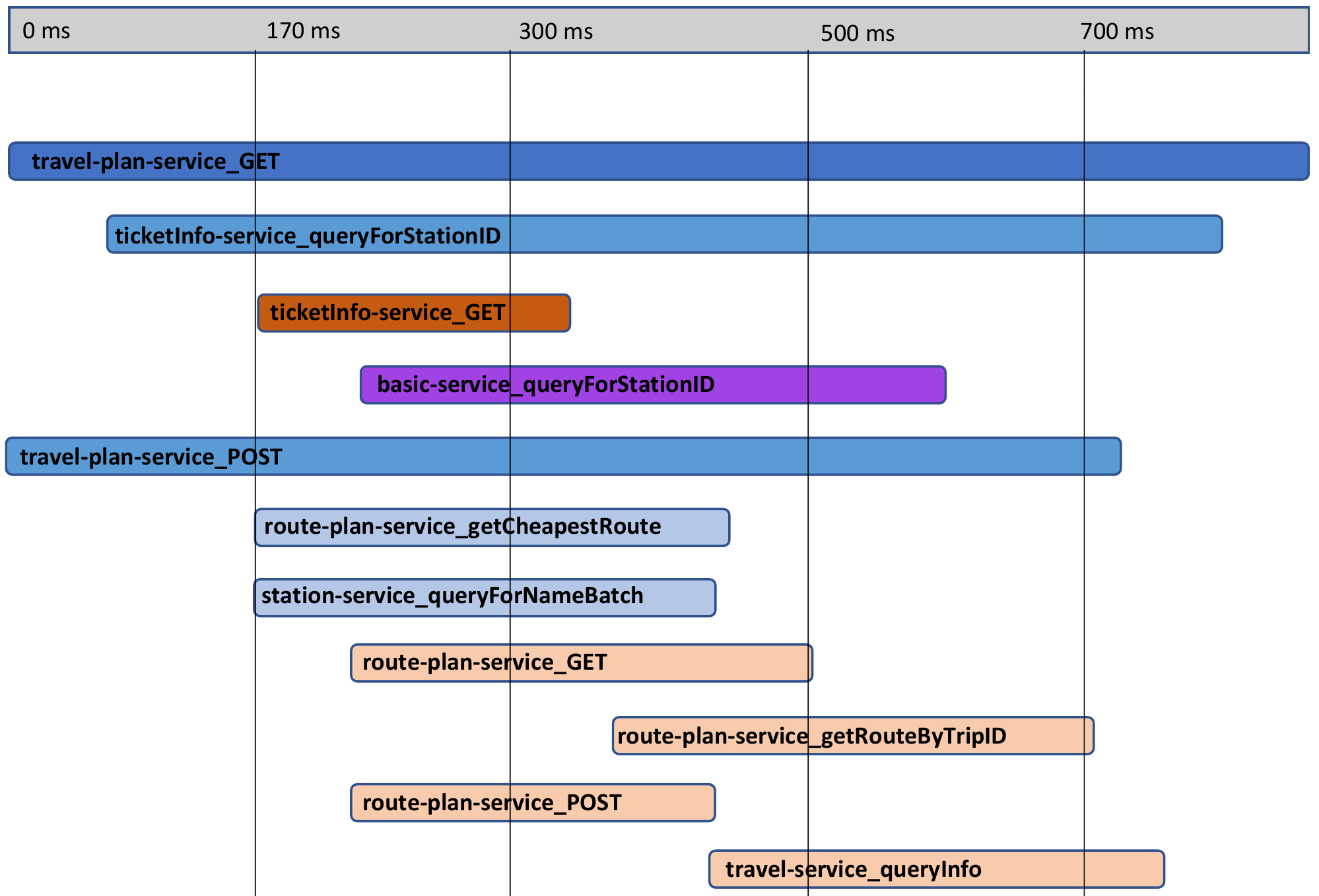}
    \caption{Swimlane visualization.}
    \label{fig:swimlane}
\end{figure}

Distributed tracing tools provide swimlane visualizations \cite{kaldor2017, sigelman2010, jaeger, davidson2022} as a means to investigate end-to-end response time (refer to \figref{fig:swimlane} for an illustration).
These visualizations are particularly useful for performance analysis of individual requests.
For example, they can be used to investigate how each RPC contributes to the end-to-end response time of a specific request, and pinpoint potential bottlenecks.

Despite their usefulness, swimlane visualizations have certain limitations when it comes to more complex performance analyses.
For instance, analyzing the performance of individual end-to-end requests can be misleading without considering the appropriate context~\cite{Davidson2023}. The response time of a request can only be considered anomalous when compared to other requests of the same type \cite{anand2020}.
Additionally, engineers are often more interested in investigating overall response time trends rather than the performance of individual requests \cite{parker2020}.
Different end-to-end response time behaviors can be associated with specific request characteristics, such as particular RPC execution paths and/or RPC performance behaviors, and engineers can be interested in identifying these characteristics to spot potential performance bugs and/or optimization opportunities~\cite{parker2020, krushevskaja2013, cortellessa2020}.

Distributed tracing tools lack support for this type of analysis, which currently requires the use of multiple visualizations and tools in tandem~\cite{Davidson2023}, \eg Jaeger~\cite{jaeger} and Kibana \cite{kibana}.
A common approach is to first recognize recurrent performance behaviors that warrant investigation, and then examine individual requests to characterize relevant performance behaviors.
This is usually achieved by detecting ``modes'' of the end-to-end response time distribution (\eg using Kibana), which indicate meaningful recurrent performance behaviors. Then, samples of requests associated with each mode can be extracted and individually examined (\eg using Jaeger's swimlanes) to identify unique characteristics that lead to specific performance behaviors, \ie modes.

Unfortunately, this approach can be particularly tedious as it necessitates the manual examination and comparison of multiple requests across multiple visualizations and tools. Furthermore, even when successful, it falls short of providing a satisfactory level of confidence. Identifying specific characteristics of a particular mode requires verifying that these characteristics appear \emph{exclusively} in requests belonging to that mode, which can be challenging using current tools. 

\begin{figure}[h]
    \centering
    \includegraphics[width=0.6\linewidth]{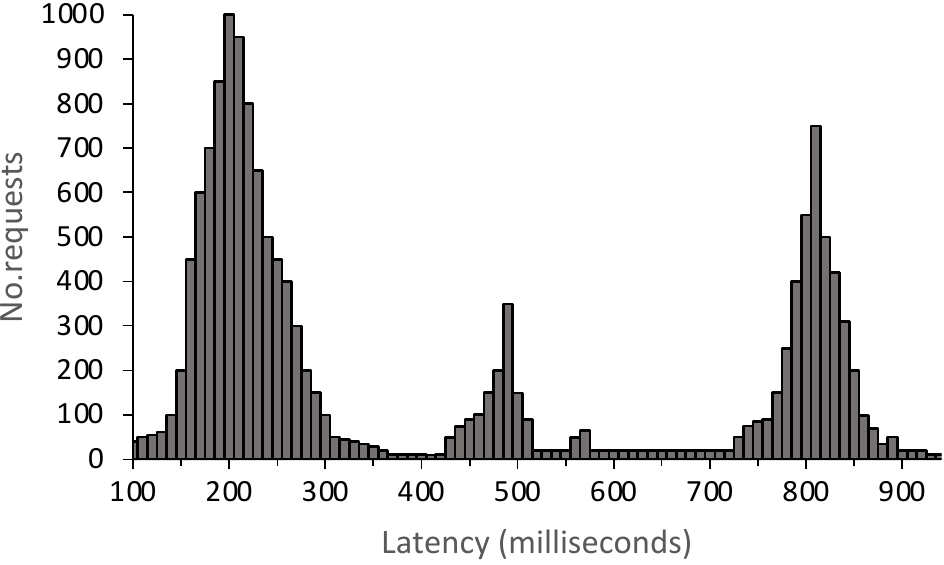}
    \caption{End-to-end response time distribution.}
    \label{fig:distribution}
\end{figure}

For instance, consider the scenario illustrated in \figref{fig:distribution}, which depicts the distribution of the end-to-end response time for a specific type of request, such as loading the homepage of a website. As depicted in the figure, requests exhibit three distinct response time behaviors, \ie modes. Suppose that the rightmost mode is characterized by a unique request characteristic, such as a specific RPC execution path. That is, a specific RPC is invoked in all requests belonging to the rightmost mode, but not in the others (\eg due to a cache miss). Identifying patterns like this can be particularly challenging with current distributed tracing tools.

\section{Proposal}

\begin{figure}[]
    \centering
    \includegraphics[width=\linewidth]{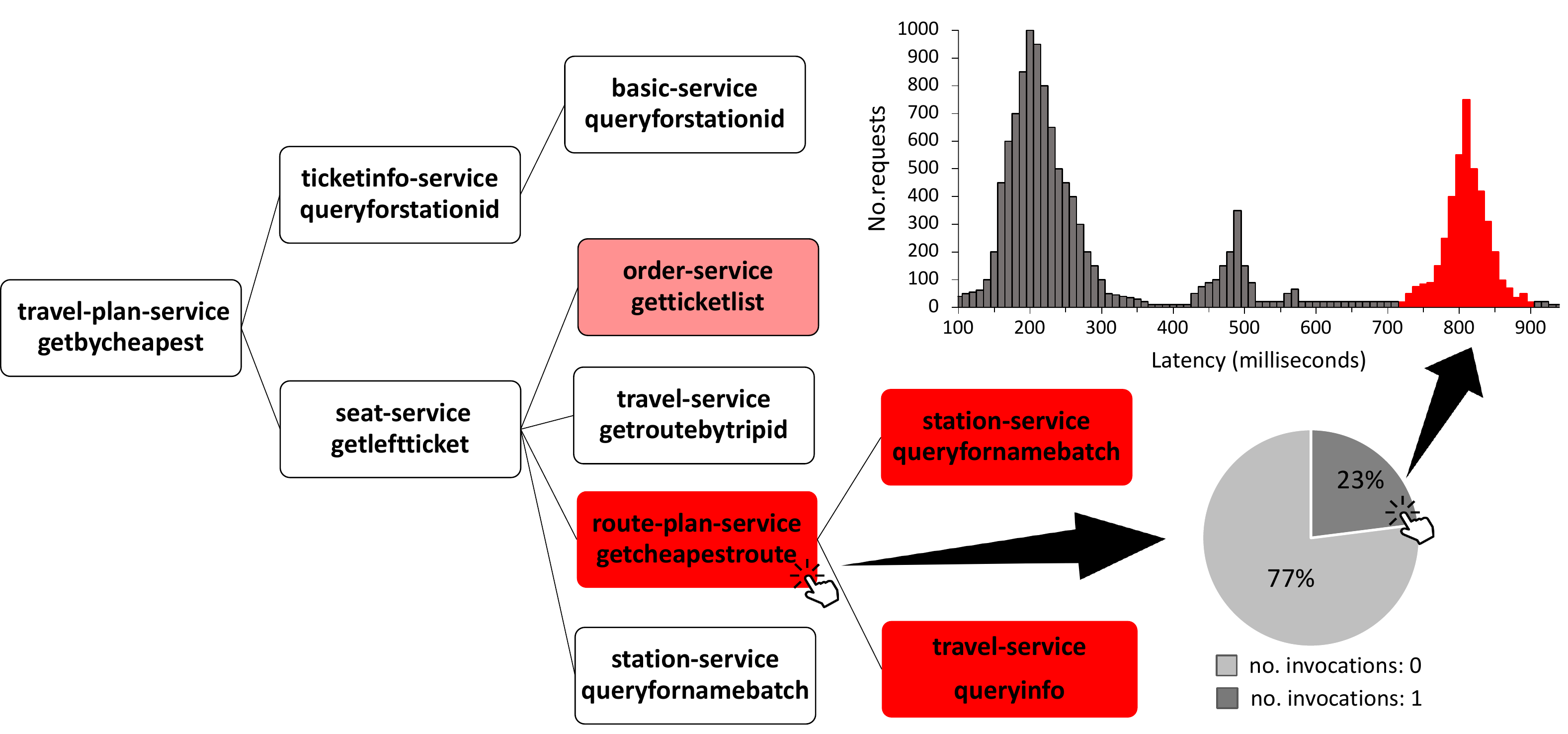}
    \caption{Dashboard.}
    \label{fig:Dashboard}
\end{figure}
We present a novel interactive visualization approach for aggregate performance analysis of end-to-end requests.
Our approach provides a dashboard comprising two main interactive components: a tree and a histogram (see \figref{fig:Dashboard}). The interactive tree offers an overview of the requests' workflows in terms of RPC execution paths, while the histogram illustrates the corresponding end-to-end response time distribution. The key insight behind this approach is to enable the simultaneous analysis of various aspects of requests and the relationship among these aspects through a unified dashboard.

The tree visualization proposed in our approach builds upon a previous visualization technique provided in Jaeger~\cite{jaeger}, which enables the comparison of two requests in terms of RPC execution paths \cite{farro2018}. We extend this technique by proposing a tree visualization that enables the analysis of an entire set of requests. Each node of the tree represents a RPC invocation within a specific execution path, where the leftmost node represents the root RPC, and edges indicate direct RPC invocation. For example, in \figref{fig:Dashboard}, the red node labeled as \texttt{getcheapestroute} indicates the RPC execution path \texttt{getbycheapest}$\rightarrow$ \texttt{getleftticket}$\rightarrow$\texttt{getcheapestroute}. It is worth noting that when a particular RPC invokes the same RPC multiple times, this leads to a single node in the tree. In other words, if the RPC \texttt{getbycheapest} invokes the RPC \texttt{getleftticket} multiple times, there will be only one child node referring to \texttt{getleftticket}. The rationale behind this approach is to address the potential complexity of requests by providing a more concise visualization. An RPC execution path will appear in the tree only if it is present in at least one request within the set of requests being analyzed.

Similar to the trace comparison tool of Jaeger~\cite{farro2018}, we utilize color encoding to highlight variations within a set of requests. However, in contrast to this approach, our color encoding scheme denotes the degree of variance of a particular request characteristic within a group of requests, rather than the difference between two requests. Furthermore, we employ a continuous color scale, as opposed to the discrete one used by Jaeger~\cite{farro2018}. As an example, we can use color encoding to depict the variability in the number of invocations of a specific RPC execution path within a group of requests. A white node indicates that the corresponding RPC execution path appears the same number of times in all the requests under analysis. Conversely, a red node indicates a substantial change in the number of invocations of the corresponding RPC execution path from one request to another. We use standardized measures of dispersion, such as the Coefficient of Variation (CV)~\cite{everitt1998}, to encode these differences into colors. For instance, a CV of 0 results in a white node, while a CV $\geq$ 1 results in a red node. The shade of color gradually transitions from white to red as the CV value increases, with 0 $\leq$ CV $\leq$ 1.

This approach facilitates the rapid identification of RPC execution paths that exhibit divergent request characteristics.
An engineer can delve deeper into a specific RPC execution path by interacting with its corresponding node, bringing up a new visualization such as the pie chart displayed in \figref{fig:Dashboard}.
This allows for a more in-depth examination of the divergent characteristics observed in that specific RPC execution path. 
The pie chart in \figref{fig:Dashboard} illustrates, for instance, two distinct behaviors, with 77\% of requests not involving any invocations of the path \texttt{getbycheapest}$\rightarrow$ \texttt{getleftticket}$\rightarrow$\texttt{getcheapestroute}, and 23\% of requests involving one invocation of such path.
 By interacting with the corresponding slice of the pie chart, the tool illustrates how this particular request characteristic affects end-to-end response time by highlighting the corresponding area of the histogram. For instance, \figref{fig:Dashboard} shows that the rightmost mode of the end-to-end response time distribution is characterized by the presence of the particular RPC execution path \texttt{getbycheapest}$\rightarrow$ \texttt{getleftticket}$\rightarrow$\texttt{getcheapestroute}.

Another advantage of this approach is its versatility in analyzing various request characteristics. This can be accomplished by changing the color encoding assigned to tree nodes. For instance, the method can be adapted to examine how different RPC execution times influence the distribution of end-to-end response time. Or, similarly, it can be utilized to investigate the impact of different HTTP header values on response time (in these situations, other statistical metrics such as the Gini index~\cite{gini1936} or Entropy~\cite{shannon1948} can be applied).
Furthermore, we plan to make the analysis bidirectional, allowing the engineer to begin their investigation from either the tree or the histogram to understand response time behavior. If the engineer chooses to start from the response time distribution, they can select a specific range using a slider selector. This selection will result in an updated color scheme on the tree, where the colors will now indicate the divergence of the selected set of requests compared to all others (using statistical measures such as Kullback-Leibler divergence~\cite{kullback1951}). This approach will enable engineers to identify how certain modes of response time distribution correlate with specific request characteristics.

\begin{figure}[]
    \centering
    \includegraphics[width=0.8\linewidth]{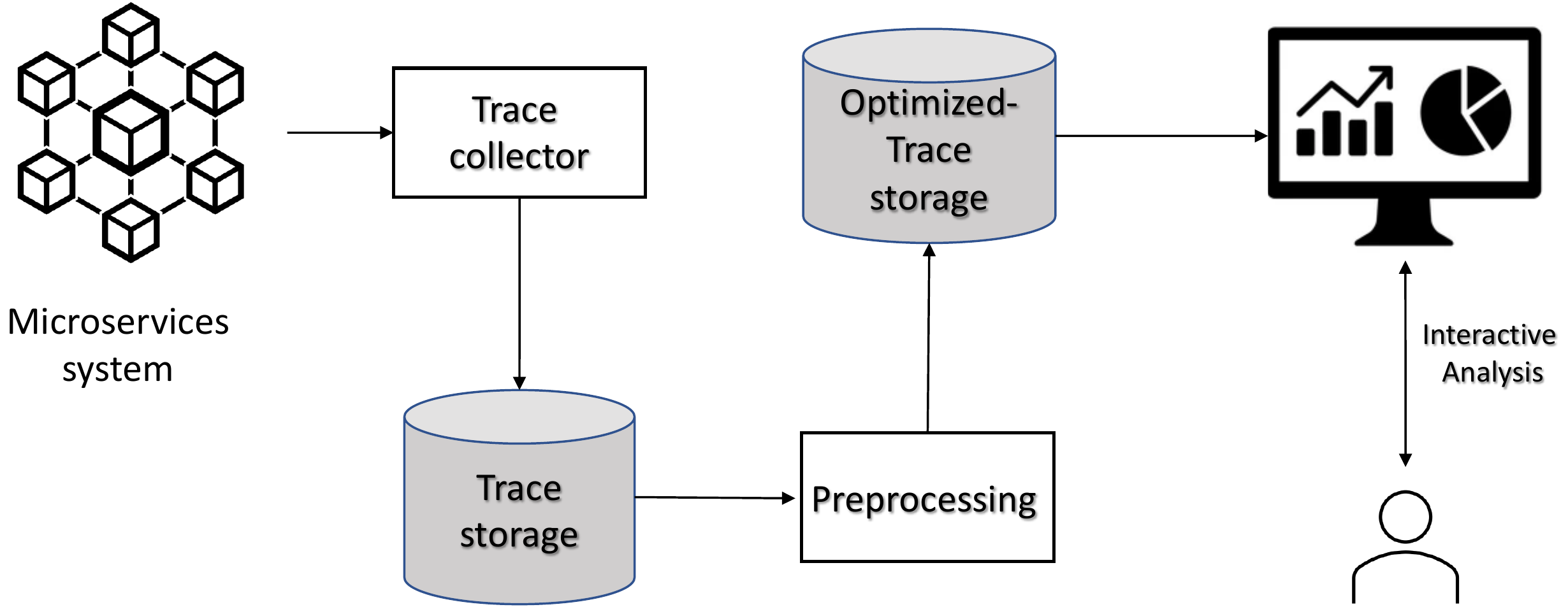}
    \caption{Workflow.}
    \label{fig:workflow}
\end{figure}

We envision two potential challenges for the development of our visualization approach.
The first challenge concerns the efficiency of our tool. Modern distributed tracing tools collect huge volumes of traces per day~\cite{kaldor2017}, which can make the interaction with the proposed visualizations computationally intensive.
In order to enable an efficient generation of visualizations, we plan to preprocess traces in an optimized format.
Figure \ref{fig:workflow} depicts the workflow of our approach. The traces of the microservices system are continuously collected by the distributed tracing tool (\ie \emph{Trace Collector}) and persisted in a \emph{Trace Storage}, such as ElasticSearch~\cite{elasticsearch}.
The stored traces are then periodically preprocessed and persisted in dedicated storage in an optimized format.
The visual dashboard application can then directly query the optimized trace storage to enable the efficient generation of visualizations.

The second challenge we envision is related to the user experience. When too much information is displayed on the screen, the user can become overwhelmed~\cite{Davidson2023}. To mitigate this, we intend to keep our dashboard as minimalistic as possible to avoid placing an excessive burden on the user. For instance, certain request types may involve a large number of RPCs, which can result in an extremely large tree structure. To reduce the user's cognitive load, we will allow the user to hide the RPCs invoked within a particular execution path by double-clicking on the corresponding node. This will enable the user to navigate the tree more easily.

\section{Ongoing work}

We are currently engaged in the development of a prototype for our visualization approach. For our pilot study, we are utilizing a dataset of traces that was generated in a previous work of the second author~\cite{traini2022b}. The dataset was created using the widely-used microservices benchmark system \emph{TrainTicket}~\cite{zhou2021,li2022,zhang2022,guo2020,zhou2019}, and the load generator \emph{PPTAM}~\cite{avritzer2019}. Our technology stack includes the use of Jaeger~\cite{jaeger} as the distributed tracing collector and Elasticsearch~\cite{elasticsearch} for storing both common and optimized traces. The dashboard has been developed using D3.js~\cite{d3} for visualizations and Flask~\cite{flask} for the backend service.

At present, we have accomplished the implementation of the UI of our working prototype, along with its key visualization components, namely the interactive tree and histogram. However, we are still engaged in the ongoing development of the preprocessing engine and a portion of the server infrastructure. For this reason, we are unable to report an authentic screenshot of our working prototype, at this stage. Nonetheless, we do provide a mock-up of the expected outcome in \figref{fig:Dashboard}.

Our next step entails the completion of the first version of our prototype.
This first version will be centered exclusively on the analysis of the number of invocations of RPC execution paths and their relationship with end-to-end performance behavior. 
  In subsequent stages, we plan to extend the capability of the prototype beyond this specific request characteristic by exploring other metrics such as RPC execution time. Additionally, we intend to enable bidirectional analyses by supporting investigations initiated from both the tree and the histogram (currently, our working prototype only supports a singular direction, \ie starting from the tree).
 
It is noteworthy that our current design only supports the individual inspection of specific request characteristics and does not enable the simultaneous analysis of multiple request characteristics in tandem.
This supplementary capability may become crucial for more elaborate performance analyses. Thus, as a long-term goal, we intend to broaden our approach by enabling the analysis of combined sets of multiple request characteristics and their interrelation with end-to-end performance.

Regarding the evaluation of our proposed approach, we plan to continue using \emph{TrainTicket} as the reference case study, given its relevance in the software engineering community. To further enhance the assessment of our proposed approach, we will consider a variety of workload scenarios, including some that involve synthetically injected performance bugs \cite{traini2022b}.
Furthermore, we aim to experiment with real-world distributed traces from large microservices systems to facilitate the transition of our approach to practice. For instance, we plan to utilize the dataset of traces recently shared by Alibaba~\cite{luo2021}.
\section{Related work}

Previous research on visualization approaches for distributed systems has traditionally focused on the analysis of individual requests or comparison between two requests. 
ShiViz \cite{beschastnikh2020} visualizes a logged distributed execution as a time-space diagram. 
The study of Sambavisan~\etal~\cite{sambasivan2013} compares three well-known visualization approaches in the context of  presenting the results of one automated performance root cause analysis approach \cite{sambasivan2011}.
TraVista \cite{anand2020} builds on the prevalent approach of visualizing a single request by augmenting this view with information to aid the user in contextualizing the performance of one request with respect to other requests. 
Jaeger~\cite{jaeger} enables the comparison of structural aspects of two requests~\cite{farro2018}.
Commercial APM tools~\cite{ahmed2016}, such as Dynatrace~\cite{dynatrace}, AppDynamics~\cite{appdynamics}, or Instana~\cite{instana}, provide features that support aggregated analysis of end-to-end requests.
For instance, Dynatrace's Service Flow feature~\cite{dynatraceflow} shows aggregate workflows of end-to-end requests along with their associated characteristics. As far as we know, current APM tools do not provide dedicated visualizations to analyze the relationship between requests' characteristics and end-to-end performance behavior. Other studies have proposed automated methods to aid the detection of patterns in requests' characteristics associated with anomalous end-to-end performance behavior~\cite{traini2022b, bansal2020, cortellessa2020, krushevskaja2013}.

Recent research by Davidson and Mace~\cite{davidson2022} emphasizes the importance of visualization in the context of systems research and encourages further effort into this topic. Davidson \etal~\cite{Davidson2023} performed a qualitative interview study to identify shortcomings of existing distributed tracing tools, and expose various open research problems that involve several domains (including visualization research).

\section{Conclusion}

In this paper, we presented a novel visualization approach for microservices performance analysis. Our approach addresses the limitations of current distributed tracing visualizations by enabling aggregate performance analysis of multiple end-to-end requests. Further studies are required to evaluate the effectiveness of our proposal and gather feedback from practitioners.

\begin{acks}
This work is supported by``Territori Aperti'' (a project funded by Fondo Territori Lavoro e Conoscenza CGIL, CSIL and UIL), and by European Union - NextGenerationEU - National Recovery and Resilience Plan (Piano Nazionale di Ripresa e Resilienza, PNRR) - Project: ``SoBigData.it - Strengthening the Italian RI for Social Mining and Big Data Analytics'' - Prot. IR0000013 - Avviso n. 3264 del 28/12/2021. We would like to thank Antinisca Di Marco and Giovanni Stilo for the useful suggestions and comments that were helpful in improving the paper.
\end{acks}

\bibliographystyle{ACM-Reference-Format}
\balance
\bibliography{references.bib}

\end{document}